\documentclass[aps,pra,showpacs,twocolumn]{revtex4}
\usepackage{graphicx}
\usepackage{bm}
\begin{document}

\title{Nonclassical states from the joint statistics of simultaneous measurements}

\author{Alfredo Luis}
\affiliation{Departamento de \'{O}ptica, Facultad de Ciencias
F\'{\i}sicas, Universidad Complutense, 28040 Madrid, Spain }

\begin{abstract}
Nonclassicality cannot be a single-observable property since  the statistics of any quantum 
observable is compatible with classical physics.  We develop a general procedure to reveal 
nonclassical behavior from the joint measurement of multiple observables. This requires 
enlarged spaces and couplings with auxiliary degrees of freedom so it must be followed by 
an inversion procedure to infer system properties from the observed statistics. In particular 
this discloses nonclassical properties of standard examples of classical-like behavior, such 
as SU(2) and Glauber coherent states. Moreover, when combined with other criteria it would 
imply that every quantum state would be nonclassical.
\end{abstract}

\pacs{03.65.Ta, 42.50.Dv, 42.50.Xa }

\maketitle

\section{Introduction}

Nonclassicality cannot be a single-observable property since within classical physics it is 
always possible to reproduce exactly the statistics of any quantum observable. For example, 
regarding optics, a classical wave with definite complex amplitude is infinitely squeezed 
and maximally subPoissionian. Moreover, we can produce classical waves whose intensity 
replicates any discrete photon number distribution as accurately as desired.  Thus, nonclassical 
effects must be found in the joint statistics of multiple observables, specially if they are incompatible 
\cite{RI}. 

Joint measurements require the coupling of the system with auxiliary degrees of freedom.
This means that in general the measurement  must  be followed by some kind of data analysis 
or inversion procedures to extract the information of the system from the observed statistics \cite{WMM}. 
We will show that when this is done in classical physics the result of the inversion is always 
a true probability distribution of the system variables. However, in quantum physics this is not 
always the case and the result of the inversion can be incompatible with classical statistics. 

This provides a very general program that includes as particular cases classic demonstrations
of nonclassical behavior, such as coincidence detection and quantum tomography. The retrieved 
system information is usually expressed in terms of the Wigner  $W$ or Glauber-Sudarshan 
${\cal P}$ functions, that actually intend to represent the joint distribution of  two complementary 
field quadratures \cite{P,QT} . 

It can be expected that this program will provide meaningful results when gathering information 
about complementary observables. But it can be as well applied successfully when addressing 
the redundant measurement of the same observable, such as photon number. Besides providing 
a new perspective on the subject, this approach discloses nonclassical properties for states with  
classical-like  ${\cal P} $ distribution, that otherwise are universally considered as classical light, 
with few exceptions  \cite{nc1,nc2,nc3}. Moreover, when combined with other nonclassical criteria 
it would imply that every quantum state would be nonclassical.

The details of the program are presented in Sec. II, including that in classical physics the result 
of the inversion is always a true probability distribution. The program is applied to a qubit 
via the measurement of complementary observables in Sec. III, showing nonclassical spin 
features for SU(2) coherent states. In Sec. IV we go beyond complementarity considering the 
redundant measurement of the photon number. Finally, in Sec. V we show a nonclassical 
effects displayed by  Glauber coherent states after the joint measurement of phase and number. 

\section{Program}
We consider the simultaneous measurement of two compatible observables $\tilde{X}$ and 
$\tilde{Y}$ which is intended to provide information about two system observables $X$ and 
$Y$ \cite{AG}. The statistics of the measurement $\tilde{p}_{X,Y} (x,y)$ is arranged so that 
each variable $x,y$ refers to the corresponding observables $X,Y$ with  marginal distributions 
\begin{equation}
\label{itn}
\tilde{p}_X (x)  = \sum_y \tilde{p}_{X,Y} (x,y), \qquad \tilde{p}_Y (y) = \sum_x
\tilde{p}_{X,Y} (x,y) .
\end{equation}
We have assumed a discrete range for  $x$ and $y$ without loss  of generality. The joint 
measurement usually requires the coupling of the system with auxiliary degrees of freedom 
that must be followed by some kind of data analysis in order to extract  information about 
$X$ and $Y$ in the system state. To this end the only hypotheses we make is that it is possible 
to infer the exact distributions $p_Z  (z)$ from their observed counterparts  $\tilde{p}_Z  (z)$, 
with $Z=X,Y$, in the form 
\begin{equation}
\label{inv}
p_Z  (z) = \sum_{z^\prime} \mu_Z (z, z^\prime ) \tilde{p}_Z  (z^\prime) ,
\quad
p_Z  = \mu_Z  \tilde{p}_Z  ,
\end{equation}
where $\mu_Z$ is a matrix  with matrix elements $\mu_Z (z, z^\prime )$ completely known 
as far as we know the measurement being performed. These are the non ideal invertible 
measurements  introduced in Ref. \cite{WMM}. 

The key idea is to extend this inversion from the marginals to the complete joint distribution 
in the form
\begin{equation}
\label{nti}
p_{X,Y} (x,y)  = \sum_{x^\prime, y^\prime} \mu_X (x, x^\prime) \mu_{Y} (y, y^\prime )  
\tilde{p}_{X,Y} (x^\prime , y^\prime ) ,
\end{equation}
or in matrix form 
\begin{equation}
\label{mf}
p_{X,Y} =\mu_X \tilde{p}_{X,Y}  \mu_Y^t ,
\end{equation}
where the superscript $t$ denotes matrix transposition. 

The retrieved distribution $p_{X,Y} (x,y)$  can be referred to as Wigner measure since by 
construction  $p_{X,Y} (x,y)$ provides the correct marginals for $X$ and $Y$ as a typical 
property of Wigner functions. In particular, when $\tilde{X}$ and $\tilde{Y}$ are field 
quadratures the result of the inversion is the classic  Wigner function \cite{WMM}. 

\bigskip
{\it Classical domain.--}
Next let us show that in the classical domain this procedure always  leads to a \textit{bona 
fide} distribution $p_{X,Y} (x,y)$.  Classically the state of the system can be completely described 
by a probability distribution $P(\lambda)$, where $\lambda$ is a point in the corresponding 
phase space. We may say that $\lambda$ represent the ontic states, while $P(\lambda)$ 
are the epistemic states \cite{OE}. The observed joint statistics is always of the form
\begin{equation}
\label{jj}
\tilde{p}_{X,Y} (x,y) = \int d \lambda P (\lambda ) \tilde{X} (x | \lambda)  \tilde{Y} (y | \lambda) ,
\end{equation}
where $\tilde{Z }(z | \lambda) $  for $Z = X,Y$ are the conditional probabilities that the 
observable $\tilde{Z}$ takes the value $z $ when the system state is $\lambda$. Imposing that 
Eq. (\ref{inv}) holds
\begin{equation}
\label{rel}
 Z( z | \lambda ) = \sum_{z^\prime} \mu_Z (z,z^\prime )  \tilde{Z} ( z^\prime | \lambda ),
\end{equation}
we readily get from  Eq. (\ref{nti}) the actual joint distribution for $X$ and $Y$
\begin{equation}
p_{X,Y} (x,y)  = \int d \lambda P (\lambda ) X(x | \lambda)  Y (y | \lambda) .
\end{equation}
This is to say that after the form (\ref{jj}) the inversion works equally well for the joint distribution 
as for the marginals. Therefore, the lack of positivity or any other pathology of the retrieved joint 
distribution $p_{X,Y} (x,y) $ is a signature of nonclassical behavior.

\section{Qubit complementarity}

The simplest example holds in a two-dimensional Hilbert space considering two complementary 
observables represented by the Pauli matrices  $X = \sigma_x$ and $Y = \sigma_y$, with 
outcomes $x,y = \pm 1$. The most general system state is represented by the density matrix in 
the system  space  $ {\cal H}_s$ 
\begin{equation}
\rho = \frac{1}{2} \left ( \sigma_0 + \bm{s}  \cdot \bm{\sigma} \right ) ,
\end{equation}
where $ \sigma_0$ is the identity, $ \bm{\sigma}$ are the three Pauli matrices, and  $\bm{s} = 
\mathrm{tr} (\rho \bm{\sigma})$ is a three-dimensional real vector with   $|\bm{s} | \leq 1$. 
The  exact statistics are
\begin{equation}
\label{mis}
p_X (x)  = \frac{1}{2} \left ( 1 +  x s_x \right ),
\qquad
p_Y (y)  = \frac{1}{2} \left ( 1 +  y s_y \right ) .
\end{equation}

For example we consider that $Y$ is measured directly  on the ${\cal H}_s$ variables by 
projection on its eigenvectors $|y = \pm 1\rangle$. The simultaneous measurement of  $X$ 
can be achieved by coupling  the system space ${\cal H}_s$ with auxiliary degrees of freedom  
${\cal H}_m$  via the following unitary operator \cite{cur}
\begin{equation}
\label{U}
U = V_+ | x=1 \rangle \langle x=1 |  +V_- |  x= -1 \rangle \langle x= -1 | ,
\end{equation}
where $V_\pm$ are unitary operators acting solely on ${\cal H}_m$ and $| x = \pm 1 
\rangle$ are the eigenvectors of $X = \sigma_x$. Denoting by $| a \rangle$  the initial state 
in ${\cal H}_m$  the transformed states are $| a_\pm \rangle = V_\pm | a \rangle \in {\cal H}_m$.
They are not orthogonal in general, and their overlap $\langle a_+ | a_- \rangle = \sin \phi$ is 
assumed to be a positive real number without loss of generality. The most simple measurement 
in ${\cal H}_m$ is described by projection on the orthonormal vectors $| \tilde{x} \rangle$
\begin{eqnarray}
\label{m12}
 | \tilde{x} = 1 \rangle &=& \frac{1}{\cos \phi} \left ( \cos \frac{\phi}{2}
\, | a_+ \rangle - \sin \frac{\phi}{2} \, | a_- \rangle \right ) ,  \nonumber \\
 | \tilde{x} = -1 \rangle &=& \frac{1}{\cos \phi} \left ( - \sin \frac{\phi}{2} \, | a_+ \rangle
+ \cos \frac{\phi}{2} \, | a_- \rangle \right ) ,
\end{eqnarray}
with $\langle 1 | a_+ \rangle = \langle -1 | a_- \rangle = \cos ( \phi/2)$, 
$\langle 1 | a_- \rangle = \langle -1 | a_+ \rangle = \sin ( \phi/2)$.

With all this, the joint statistics for the simultaneous measurement is 
\begin{equation}
\label{js}
\tilde{p}_{X,Y} (x,y) = \frac{1}{4} \left ( 1 + x s_x \cos \phi   + y s_y \sin \phi \right  ) ,
\end{equation}
with marginals 
\begin{equation}
\label{ms}
\tilde{p}_X  (x) = \frac{1}{2} \left ( 1 +  x s_x \cos \phi  \right ) ,
\qquad
\tilde{p}_Y (y)  = \frac{1}{2} \left ( 1 +  y s_y \sin \phi   \right )  ,
\end{equation}
which can be easily inverted in the form
\begin{equation}
 p_Z = \mu_Z  \tilde{p}_Z, \qquad 
\mu_Z \left ( z,  z^\prime \right ) = \frac{1}{2} \left ( 1 + \frac{z z^\prime }{\eta_Z } \right ) , 
\end{equation}
with  $\eta_X = \cos \phi$,  $\eta_Y = \sin \phi$, and $z,z^\prime = \pm1$. Finally, the 
inversion applied to the joint statistics $p_{X,Y} = \mu_X \tilde{p}_{X,Y}  \mu^t_Y$ leads to  
\begin{equation}
p_{X,Y} (x, y) = \frac{1}{4} \left ( 1 + x s_x + y s_y  \right  ) .
\end{equation}
This is not a true Wigner function in the sense that the information about the $\sigma_z$ 
observable is absent.

Clearly $p_{X,Y} (x,y)$ can take  negative values if $|\bm{s} | > 1/\sqrt{2}$. The states presenting 
the largest negativity are pure states with $| s_x |= |s_y | = 1 /\sqrt{2}$  and $s_z =0$. The system 
space for the qubit is the unit Bloch sphere. Relying on the spherical symmetry we may say that 
any state with $|\bm{s}| > 1/ \sqrt{2}$ is actually nonclassical after a suitable choice of the $X$, 
$Y$ observables. This is that the 65 \% of the volume of the Bloch sphere is occupied by nonclassical 
states. This result is quite relevant since it is often argued that all states in a two-dimensional space 
are classical regarding spin variables, as far as all them have a well-behaved Glauber-Sudarshan 
SU(2) ${\cal P}$ function \cite{GBB}. Actually, all the pure states are SU(2) coherent states and have 
the maximum negativity.

\section{Number-number measurements}

Next the system is an electromagnetic field mode. This is mixed with vacuum at a lossless 50\% 
beam splitter. The observables to be measured $\tilde{X}$, $\tilde{Y}$ are the number of photons 
$n_1$ and $n_2$ registered at the two output ports. Both are intended to provide information 
about the number of photons in the system state $\rho$, this is $X = Y = n$. The observed joint 
statistics is (for simplicity we drop the subscripts in $p$ and $\tilde{p}$ since there is no risk of 
confusion) 
\begin{equation}
\tilde{p} (n_1, n_2) = \frac{1}{2^{n_1+n_2}} \pmatrix{ n_1+ n_2 \cr n_1} p (n_1+ n_2 ) ,
\end{equation}
with identical marginals for both variables $n = n_1,n_2$
\begin{equation}
\label{direct}
\tilde{p} (n) = \sum_{k=n}^\infty \frac{1}{2^k} \pmatrix{ k \cr n} p (k ) ,
\end{equation}
where $p(n) = \langle n | \rho | n \rangle$ is the true number statistics of the system state.
The inversion of the marginals can be carried out via a transformation readily similar 
to Eq. (\ref{direct})  
\begin{equation}
 p (n ) = (-2)^n \sum_{k=n}^\infty  \pmatrix{ k \cr n}  (-1)^k  \tilde{p} (k)  .
\end{equation}
When the inversion is applied to the two variables in the joint distribution $\tilde{p} (n_1, n_2)$
via Eq. (\ref{nti}) we get a distribution $p(n_1, n_2)$ that after a little algebra can be expressed as 
\begin{equation}
 p (n_1 , n_2  ) =  \sum_{k=0}^\infty \frac{(n_1 + n_2 +k)!}{n_1 ! n_2 ! k!}  (-1)^k 
 p ( n_1+ n_2 + k )  ,
\end{equation}
with a clear symmetry between $n_1$ and $n_2$. This adopts a quite simple expression if we use 
the Glauber-Sudarshan distribution ${\cal P} (\alpha)$ in the form 
\begin{equation}
p (n_1 , n_2  ) = \frac{1}{n_1 ! n_2 !} \int d^2 \alpha {\cal P} (\alpha) | \alpha |^{2n_1+2n_2} 
e^{- 2 | \alpha |^2} .
\end{equation}
In particular for $n_1 = n_2 = 0$ we get that $ p (0,0  )$ is the Wigner function $W(\alpha)$ at the 
origin $\alpha =0$ as proportional to the mean value of the parity operator
\begin{equation}
 p (0,0  ) = \frac{\pi}{2} W ( \alpha = 0 ).
\end{equation}
From the above relations  we get that if ${\cal P} (\alpha)$ is a classical-like distribution so will be 
$ p (n_1 , n_2  )$, provided it exists.

\bigskip

{\it Number states.--}
A readily example is provided by the photon-number states $| n \rangle$. For example for the  
one-photon state $n = 1$ we have $p (0,0) = -1$, $p (1,0) = p(0,1) = 1$, while all the other 
vanish $p(n_1\neq 0 , n_2 \neq 0 ) = 0$. For states with larger number of photons  we get results 
as illustrated by the chessboard in Fig. 1 for $| n = 7 \rangle$. The  blue squares represent  
negative values for $p (n_1 , n_2  )$, the darker the more negative, with an absolute minimum at 
$n_1 = n_2 = 2$ with $p( 2, 2 ) = -210$.  Actually, the absolute value of the most negative 
$p (n_1 , n_2  )$ increases heavily with the number of photons $n$, with rather strong 
positive-negative oscillations.  

\begin{figure}
\begin{center}
\includegraphics[scale=0.5]{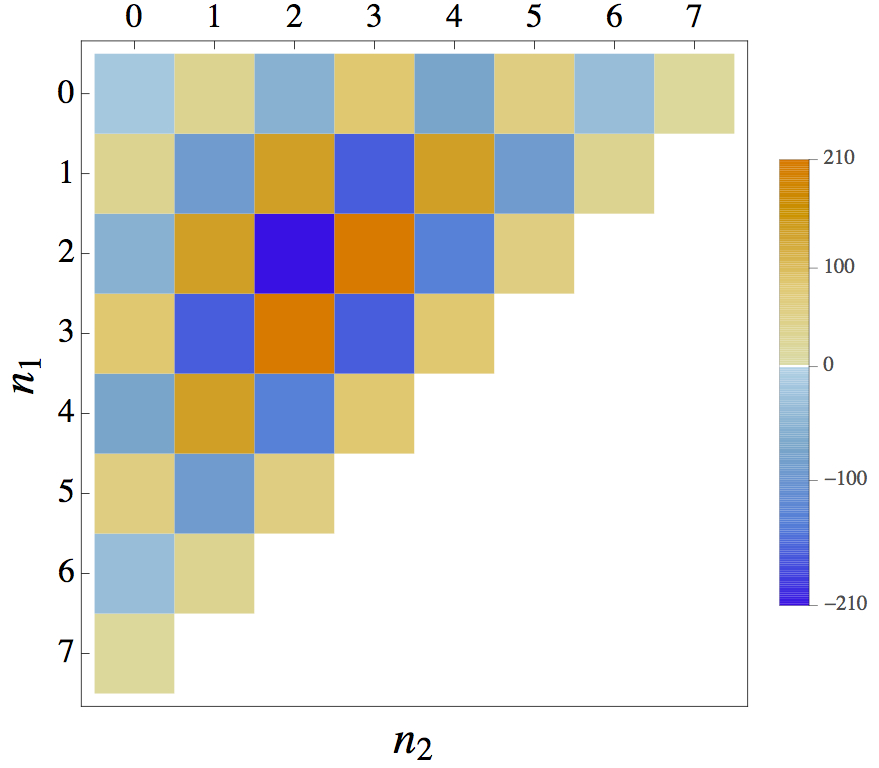}
\end{center}
\caption{Chessboard with the $p (n_1 , n_2  )$ values for a number state  $| n = 7 \rangle$. The  
blue squares represent negative $p (n_1 , n_2  )$ values, the darker the more negative. }
\end{figure}

\bigskip

{\it Photon added thermal states.--}
As a more realistic version of the number states we may consider the photon-added thermal states 
$\rho \propto a^{\dagger k} \rho_t a^k$ \cite{PATS}, where $\rho_t$ is a thermal state, with photon-number 
distribution
\begin{equation}
p(n \geq k) = \frac{1}{(\bar{n}+1)^{k+1}} \pmatrix{n \cr k} \left ( \frac{\bar{n}}{\bar{n}+ 1} \right )^{n-k} ,
\end{equation}
and $p(n < k) =0$, where $\bar{n}$ is the mean number of photons of the thermal state. It can be easily 
seen that for the single-photon-added thermal state $k=1$ the only negative value $p(n_1,n_2)  < 0$ 
holds for $n_1 = n_2 =0$ with $p(0,0) = - 1/(2\bar{n} + 1)^2$, as illustrated in Fig. 2.

\begin{figure}
\begin{center}
\includegraphics[scale=0.5]{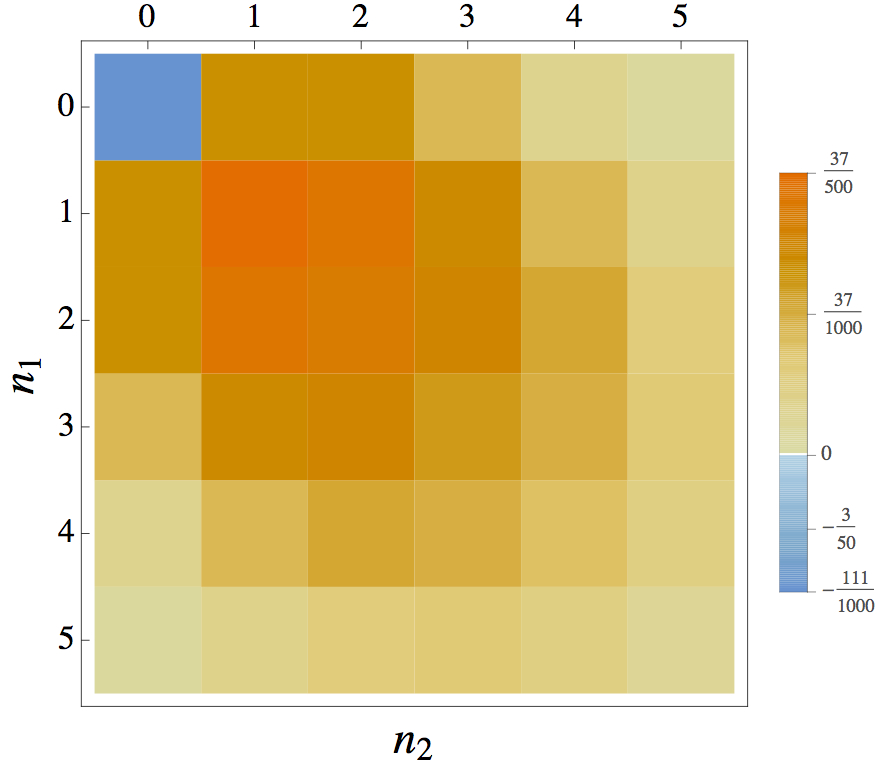}
\end{center}
\caption{Chessboard with the $p (n_1 , n_2  )$ values for a single-photon-added thermal state  
$k=\bar{n}=1$. The  blue squares represent negative $p (n_1 , n_2  )$ values, the darker the more 
negative. }
\end{figure}

\bigskip

{\it Squeezed vacuum.--}
As a further example we may consider the squeezed vacuum state  with photon-number distribution that 
contains only  contributions for even numbers \cite{SV}
\begin{equation}
p(n ) = \frac{n!}{2^n (n/2)!^2} \frac{\tanh^n r}{\cosh r} ,
\end{equation}
with $p(n)=0$ for odd $n$, where $r$ is the squeezing parameter and $\bar{n} = \sinh^2 r$ is the mean 
number of photons. The most  negative result $p(n_1,n_2) < 0$ holds always for  $p(0,1) =  p(1,0)= -  
\bar{n}$ as illustrated in Fig. 3 for $r=0.3$ and $\bar{n}\simeq 0.1 $. Numerically we have shown that the 
inversion for the joint distribution works only for $\bar{n} \leq 0.1$, while for the marginals it still works 
about this value. 
\begin{figure}
\begin{center}
\includegraphics[scale=0.5]{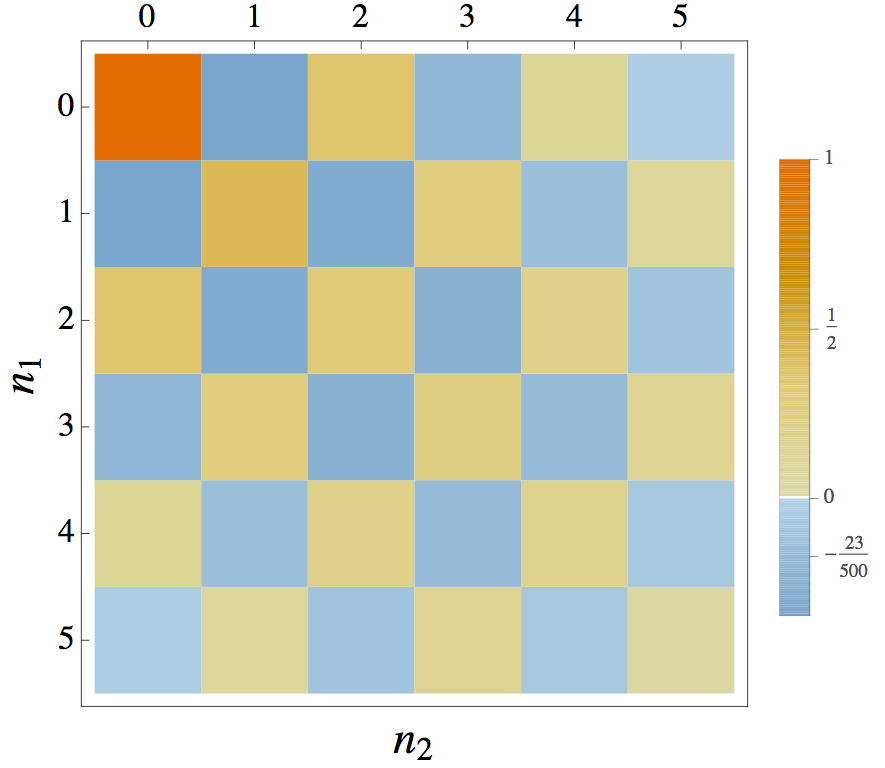}
\end{center}
\caption{Chessboard with the $p (n_1 , n_2  )$ values for a squeezed vacuum state with $\bar{n} \simeq 0.1$. 
The  blue squares represent negative $p (n_1 , n_2  )$ values, the darker the more negative . }
\end{figure}

\section{Phase-number measurements}

Let us address again a measurement scheme with two photon-number detectors placed at the 
output ports  of a 50 \% lossless beam splitter providing a joint statistics $\tilde{p}^\prime (n_1, n_2)$. 
However, now the beam splitter mixes  the system state $\rho$ with a Glauber coherent state 
$| \beta \rangle$. In this case, the variables of interest  are not the number of photons recorded in 
each detector  $n_{1,2}$,  but the total number $N= n_1 + n_2$ and the normalized number difference 
$m=(n_1-n_2)/(n_1 + n_2)$ with $m \in [-1,1]$. The measured joint distribution for  these variables is 
\begin{equation}
\tilde{p} (N, m) = \tilde{p}^\prime \left ( n_1= N \frac{1+m}{2}, n_2 = N \frac{1- m}{2}  \right ) ,
\end{equation}
where $N=0, 1, \ldots, \infty$, while the range for $m$ is the union of all the ranges allowed
for each $N$, this is $m = -1, -1+2/N,\ldots, 1$, and we consider just $m=0$ for $N=0$. 

The marginal statistics for the total number $\tilde{p}(N)$ can be regarded as providing information  
about the photon-number distribution $p(n) = \langle n | \rho | n \rangle$ of  the system state $\rho$. 
On the other hand, the marginal distribution $\tilde{p}(m)$ is a suitable version of the cosine of the 
relative phase.  For the sake of simplicity we consider that $\tilde{p} (m )$ requires no further data 
transformation. This is consistent with the understanding that there is no operator for the single-mode 
phase, and that this is a somewhat fuzzy variable represented by positive operator measures 
\cite{pp,qrp}. In particular, the variable $m$ bears some resemblance with relative-phase approaches 
introduced via homodyne detection \cite{RL,AC,cps}.

The marginal for the photon-number sum is 
\begin{equation}
\label{Nn}
\tilde{p} (N) = \sum_{m \in M} \tilde{p} (N, m) = 
e^{-\bar{n}} \sum_{n=0}^N  \frac{\bar{n}^{N-n}}{(N-n)!} p(n)  ,
\end{equation}
where $\bar{n} = | \beta |^2$ is the mean number of photons of the coherent state $| \beta \rangle$. 
The above relation (\ref{Nn}) can be inverted in the form 
\begin{equation}
\label{nN}
p (n) = e^{\bar{n}} \sum_{N=0}^n  \frac{(-\bar{n})^{n-N}}{(n-N)!}  \tilde{p} (N)  ,
\end{equation}
which is essentially the same transformation (\ref{Nn}) after replacing $\bar{n}$ by $-\bar{n}$.  The 
program ends applying the inversion transformation to the complete joint distribution 
\begin{equation}
\label{jNm}
p (n,m) = e^{\bar{n}}  \sum_{N=0}^n \frac{\left (-\bar{n} \right )^{n-N}}{(n-N)!} \tilde{p} (N,m)  .
\end{equation}

A readily demonstration that this can disclose nonclassical behavior holds when considering  the 
particular values $n=1$ and $m=0$, taking into account that $m=0$ is not included for $N=1$
\begin{equation}
\label{pc}
p (1, 0) = - \bar{n} e^{\bar{n}}   \tilde{p} (0,0) = - \bar{n} e^{\bar{n}}    \tilde{p}^\prime (0, 0) 
=    - \bar{n}  p( 0) . 
\end{equation}
Therefore, whenever $p( 0) \neq 0$ and $\bar{n} \neq 0$ we have  $p (1,0) < 0$. 

This result may be ascribed to same kind of effective entanglement between the $N$ and $m$ 
variables since the allowed $m$ may be different for different $N$. This is a purely quantum-phase 
effect since in the classical case there is no relation whatsoever between the intensity and the 
relative  phase that can take any value between -1 and 1 for any value of the intensity. This is 
reflected by most approaches to the quantum relative phase \cite{qrp}. Equivalent results are 
obtained for other definitions of the $m$ variable such as $m=n_1 - n_2$ or even $m=n_1$. 

\bigskip

{\it Nonclassicality of Glauber coherent states.--}
In particular Eq. (\ref{pc}) discloses non classical properties for the Glauber coherent states $| \alpha \rangle$, 
which are widely assumed as classical-like states. The most simple example is the vacuum state $| \alpha = 0 
\rangle$ with  
\begin{equation}
\tilde{p}^\prime \left ( n_1, n_2 \right )  = \frac{\left ( \bar{n}/2 \right )^{n_1+n_2}}{n_1 ! n_2 !} e^{-\bar{n}} ,
\end{equation}
and 
\begin{equation}
\tilde{p} (N,m )  = \frac{\left ( \bar{n}/2 \right )^N}{\frac{N(1+m)}{2} ! \frac{N(1- m)}{2} !} e^{-\bar{n}} .
\end{equation}
The result of the inversion (\ref{jNm}) is illustrated in Fig. 4 for $\bar{n}=1$ where the most negative value 
is $P(N=1,m=0)=-1$.

The fact the coherent states display nonclassical behavior for number-phase variables was  already noticed in 
Refs. \cite{nc1,nc2}.

\begin{figure}
\begin{center}
\includegraphics[scale=0.5]{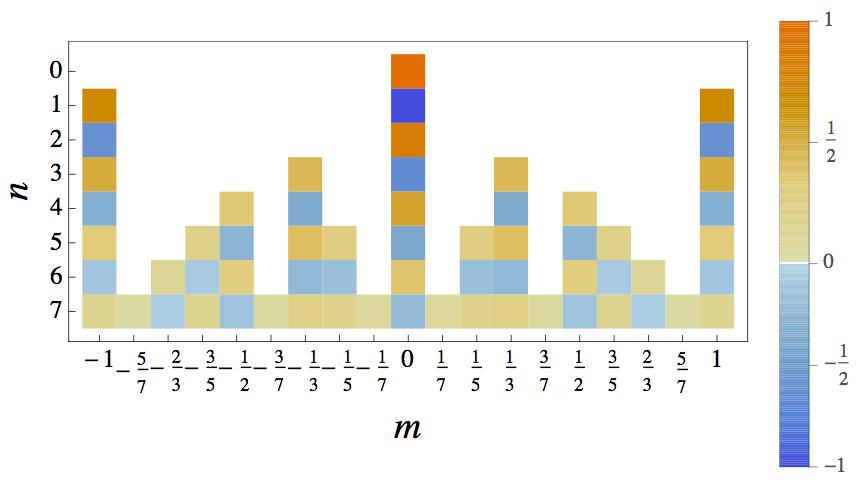}
\end{center}
\caption{Chessboard with the $p (N , m  )$ values for $\rho$ in the vacuum state mixed in a 50 \% beam splitter 
with a reference coherent state with $\bar{n}=1$. The  blue squares represent negative $p (N , m )$ values, the 
darker the more negative. }
\end{figure}

\bigskip

{\it Every quantum state is nonclassical .--}
To some extent the result in Eq.  (\ref{pc}) is the dual of the nonclassical Lee criterion in Refs. \cite{Lee}, where 
maximal non classicality is obtained by the lack of vacuum contribution. In other words, the union of both criteria 
would imply that all quantum states are nonclassical. 

\bigskip

\section{Conclusions}

Summarizing, we have addressed an universal protocol to disclose nonclassical behavior via joint measurements 
of more than one observable. This includes previous criteria and admits many other possibilities 
under one and the same framework. In particular this reveals nonclassical properties for standard examples of 
classical-like behavior, such as SU(2) and Glauber coherent states. Moreover, when combined with other 
criteria implies that every quantum state is nonclassical.

\section*{ACKNOWLEDGMENTS}

Support is acknowledged from project FIS2012-35583 of Ministerio de Econom\'{\i}a y Competitividad and 
the CAM research consortium QUITEMAD+ S2013/ICE-2801.


\begin{thebibliography}{00}

\bibitem{RI}
A. Rivas, arXiv:1501.04929 [quant-ph].

\bibitem{WMM}
W. M. Muynck, \textit{Foundations of Quantum Mechanics, an Empiricist Approach}, 
(Kluwer Academic Publishers, Dordrecht, The Netherlands, 2002);
W. M. de Muynck and H. Martens, Phys. Rev. A \textbf{42}, 5079 (1990);
W. M. de Muynck, Phys. Lett. A \textbf{182}, 201 (1993); 
J. Phys. A: Math. Gen. \textbf{31}, 431(1998).

\bibitem{P} 
L. Mandel and E. Wolf, \textit{Optical Coherence and  Quantum Optics} 
(Cambridge University Press, Cambridge, England, 1995);
V. V. Dodonov, J. Opt. B: Quantum Semiclass. Opt. \textbf{4}, R1 (2002).

\bibitem{QT} 
W. Vogel and H. Risken, Phys. Rev. A \textbf{40}, 2847 (1989);
D.-G. Welsch, W. Vogel, and T. Opatrn\'{y}, in  \textit{Progress in Optics}, edited 
by E. Wolf (Elsevier Science, Amsterdam, 1999), Vol. 39.

\bibitem{nc1}
J. Vaccaro,  Phys. Rev. A {\bf 52},  3474  (1995).

\bibitem{nc2}
A. Luis,  Phys. Rev. A {\bf 73},  063806 (2006).

\bibitem{nc3}
L. M. Johansen, Phys. Lett. A  {\bf 329}, 184 (2004).

\bibitem{AG}
K. W\'{o}dkiewicz, Phys. Rev. Lett. {\bf 52}, 1064  (1984); Phys. Lett. A {\bf 124}, 207 (1987);
E. Arthurs and M. S. Goodman, Phys. Rev. Lett. {\bf 60}, 2447 (1988);
S. Stenholm, Ann. Phys. (N.Y.) {\bf 218}, 233 (1992);
M. G. Raymer, Am. J. Phys. {\bf 62}, 986 (1994);
W. M. de Muynck, Found. Phys. {\bf 30}, 205 (2000);
T. Brougham, E. Andersson and S.M. Barnett, Phys. Rev. A \textbf{80}, 042106 (2009).

\bibitem{OE} 
R.  W. Spekkens, Phys. Rev. A \textbf{71}, 052108 (2005); Phys. Rev. Lett.  \textbf{101}, 020401 (2008);
M. J.W. Hall, 0909.0015v3 [quant-ph]; 
N. Harrigan and R.  W. Spekkens, Found. Phys \textbf{40}, 125 (2010);
M. Zukowski and C. Brukner, J. Phys. A: Math. Theor. {\bf 47}, 424009 (2014).
 
\bibitem{cur}
A. Luis, Phys. Rev. A \textbf{84}, 034101 (2011);  arXiv:1306.5211 [quant-ph];
A. Luis, G. M.  Bosyk, and M. Portesi, arXiv:1501.06667 [quant-ph].

\bibitem{GBB}
F. T. Arecchi, E. Courtens, R. Gilmore, and H. Thomas,  Phys. Rev. A \textbf{6}, 2211 (1972);
O. Giraud, P. Braun, and D. Braun,  Phys. Rev. A \textbf{78}, 042112 (2008);
New J.  Phys. \textbf{12},  063005 (2010).

\bibitem{PATS} 
G. S. Agarwal and K. Tara, Phys. Rev. A 46, 485 (1992);
T. Kiesel, W. Vogel, V. Parigi,  A. Zavatta and M. Bellini, Phys. Rev. A \textbf{78}, 021804 (2008);
T. Kiesel, W. Vogel, M. Bellini and A. Zavatta, Phys. Rev. A \textbf{83}, 032116 (2011).

\bibitem{SV} 
M. O. Scully and M. S. Zubairy, \textit{Quantum Optics}
(Cambridge University Press, Cambridge, England, 1997).

\bibitem{pp} 
M. Grabowski, Int. J. Theor. Phys. {\bf 28}, 1215 (1989); Rep. Math. Phys. {\bf 29}, 377 (1991);
J. Bergou  and B.-G. Englert, Ann. Phys. (N. Y.) {\bf 209}, 479 (1991);
R. Lynch, Phys. Rep. {\bf 256}, 368 (1995);
V. Pe\v{r}inov\'{a},  A. Luk\v{s} and J. Pe\v{r}ina, {\em Phase in Optics}  (World Scientific, Singapore, 1998).

\bibitem{qrp}
A. Luis and L. L. S\'{a}nchez-Soto, \textit{Progress in Optics}, edited by E. 
Wolf (Elsevier, Amsterdam, 2000) {\bf 41},  421 (2000);
Phys. Rev. A \textbf{48},  4702 (1993). 

\bibitem{RL}
R. Lynch, J. Opt. Soc. Am. B \textbf{4}, 1723 (1987).

\bibitem{AC}
A. Cives-Esclop, A. Luis, and L. L. S\'{a}nchez-Soto, Opt. Commun. \textbf{175}, 153 (2000).

\bibitem{cps} 
P. Carruthers and M. M. Nieto, Rev. Mod. Phys. {\bf 40}, 411 (1968);
S. M. Barnett and D. T. Pegg, J. Phys. A {\bf 19}, 3849 (1986);
J. W. Noh, A. Foug\`{e}res and L. Mandel, Phys. Rev. Lett. {\bf 67}, 1426 (1991);
Phys. Rev. A {\bf 45}, 424 (1992); Phys. Rev. A {\bf 46}, 2840 (1992);
A. S. Shumovsky, Opt. Commun. {\bf 136}, 219 (1997).

\bibitem{Lee} 
C. T. Lee, Phys. Rev. A \textbf{44}, R2775 (1991); \textbf{52}, 3374 (1995);
N. L\"{u}kenhaus and S. M. Barnett, \textit{ibid.} \textbf{51}, 3340 (1995);
A. F. de Lima and B. Baseia, \textit{ibid.} \textbf{54}, 4589 (1996);
J. Janszky, M. G. Kim, and M. S. Kim, \textit{ibid.} \textbf{53}, 502 (1996).



\end{thebibliography}
\end{document}